\documentclass[pdflatex,sn-mathphys-num]{sn-jnl}

\usepackage{graphicx}%
\usepackage{multirow}%
\usepackage{amsmath,amssymb,amsfonts}%
\usepackage{amsthm}%
\usepackage{mathrsfs}%
\usepackage[title]{appendix}%
\usepackage{xcolor}%
\usepackage{textcomp}%
\usepackage{manyfoot}%
\usepackage{booktabs}%
\usepackage{algorithm}%
\usepackage{algorithmicx}%
\usepackage{algpseudocode}%
\usepackage{listings}%
\usepackage{placeins}

\theoremstyle{thmstyleone}%
\theoremstyle{thmstyletwo}%

\theoremstyle{thmstylethree}%

\begin{document}

\title[Article Title]{Toward an Origin of Human Randomness: Interaction-Driven Enhancement in the Rock--Paper--Scissors Game}

\author*[1]{\fnm{Song-Ju} \sur{Kim}}\email{kim@sobin.org}

\author[2]{\fnm{Shoma} \sur{Ohara}}\email{ohara.shoma@nit.ac.jp}

\author[3]{\fnm{Hiroaki} \sur{Kurokawa}}\email{hkuro@stf.teu.ac.jp}

\affil*[1]{\orgname{SOBIN Institute LLC}, \orgaddress{\street{3-38-7 Keyakizaka}, \city{Kawanishi}, \postcode{666-0145}, \state{Hyogo}, \country{Japan}}}

\affil[2]{\orgdiv{Faculty of Fundamental Engineering}, \orgname{Nippon Institute of Technology}, \orgaddress{\street{4-1 Gakuendai, Miyashiro}, \city{Minamisaitama}, \postcode{345-0826}, \state{Saitama}, \country{Japan}}}

\affil[3]{\orgdiv{School of Engineering}, \orgname{Tokyo University of Technology}, \orgaddress{\street{1404-1 Katakura}, \city{Hachioji}, \postcode{192-0982}, \state{Tokyo}, \country{Japan}}}


\abstract{
 Human-generated randomness is constrained by cognitive, motor, and strategic biases. This study examines how these constraints appear in individual behavior and how they may be modified through interaction with another human. We analyzed repeated rock--paper--scissors data from 9 participants, yielding 108 human--human matches and 216 individual player sequences.

 Using Lempel--Ziv complexity (LZC), we compared human--human sequences with the RNG-opponent condition. In the RNG-opponent condition, the maximum human LZC value was 84, which we used as an empirical reference. In the human--human condition, most sequences remained below this value, but a small number exceeded it, producing a small high-complexity tail that was not present in the RNG-opponent condition.

We introduced a sensitivity measure that captures whether a player responds to the opponent's recent frequency bias by choosing the move that beats the opponent's most frequent recent move. Partial regression showed that focal-player sensitivity positively predicted future entropy in the opponent's move sequence after controlling for the opponent's current entropy. Circular-shift surrogate analyses indicated that this relation was most clearly interaction-specific when the opponent was in a low-entropy state, where the recent move distribution contained a clear frequency bias.

These results suggest that human randomness is not only an isolated individual capacity, but can be shaped by interaction in a state-dependent manner.
The findings identify a local mechanism by which interaction may destabilize biased behavior and increase entropy, providing a concrete basis for future causal experiments and generative models of high-complexity human behavior.
}



\maketitle

\section{Introduction}
\label{sec:introduction}

Can interaction between intelligent systems help them escape their own regularities?
Human behavior is generated by a highly organized biological system, and this organization supports stable perception, prediction, and action.
At the same time, the same organization may constrain the ability of an isolated individual to generate irregular or random-like behavior.
This paper examines whether interaction with another human can disturb such individual regularities and enhance operational measures of randomness.

Randomness is a basic concept in science, technology, and human culture. It is used in probability theory, statistics, cryptography, simulation, games, decision making, and the study of complex systems. At the same time, randomness has different meanings depending on the context. In everyday language, randomness is often associated with unpredictability or lack of control. In mathematics, however, randomness is defined more carefully. Martin-L{\"o}f randomness defines randomness of infinite sequences in terms of passing all effectively specified statistical tests \cite{MartinLof1966}. In computational pseudorandomness, Yao's next-bit test theorem gives a related but complexity-theoretic characterization, connecting next-bit unpredictability with indistinguishability by polynomial-time statistical tests \cite{Yao1982}. Related ideas are developed in algorithmic information theory, where randomness is connected to incompressibility and Kolmogorov complexity \cite{LiVitanyi2008}. In practical applications, however, one usually works with finite sequences and finite batteries of tests, such as the NIST statistical test suite \cite{Rukhin2010NIST,KimUmenoHasegawa2003NIST}.

This distinction is important for the present study. For a finite behavioral sequence, one cannot prove absolute randomness in the Martin-L{\"o}f sense or test indistinguishability against all possible procedures. One can only evaluate whether particular forms of non-randomness are detected by the tests or measures being used. Thus, in this paper, randomness is treated operationally. We do not ask whether human rock--paper--scissors sequences are truly random in an absolute mathematical sense. Instead, we ask whether interaction between two humans can increase operational measures of randomness and complexity, especially Lempel--Ziv complexity and entropy \cite{LempelZiv1976}.

This issue becomes especially important when randomness is generated by humans. Human-generated randomness is not simply an imperfect version of a mathematical random sequence. It is a behavioral product of a biological system with neural, motor, cognitive, and strategic constraints. A long line of psychological research has shown that humans often fail to generate statistically random sequences when instructed to do so. Their sequences tend to contain biases such as too many alternations, too few long runs, local representativeness, or stereotyped response patterns \cite{Wagenaar1972,Nickerson2002,RapoportBudescu1997}. Random sequence generation has also been linked to feedback and executive control of working memory \cite{Neuringer1986,Baddeley1998}.

These limitations can be understood as constraints on the effective search space of human behavior. Even when a person intends to act randomly, the actual sequence of actions may remain confined to a relatively small set of habitual or cognitively available patterns. In this sense, the problem of human randomness is not only that humans fail to imitate ideal random sequences. Rather, the problem is that an organized biological system may have difficulty escaping its own internal regularities. The same structure that supports stable perception, memory, prediction, and action may also make it difficult for an isolated individual to generate fully irregular behavior.

Historically, humans have often used physical devices to generate randomness. Dice, coins, cards, and other physical randomizers have been used in games, gambling, divination, and decision making. The history of gambling and random devices is closely related to the emergence of probability as a mathematical and scientific concept \cite{Hacking1975EmergenceProbability}. This historical fact is suggestive. When humans wanted randomness, they often did not rely only on their own internal decision processes. They used external physical systems that were regarded as difficult to control or predict. One possible interpretation is that humans implicitly recognized a limitation in internally generated randomness and used physical devices to escape from their own cognitive and behavioral constraints.

Interaction with another human may change this situation. In strategic interaction, another person is not a passive random source, but a dynamic environment that detects, exploits, and responds to one's biases. If a player tends to repeat a move, alternate too regularly, or favor one action, the opponent may use this bias strategically. The player is then forced to update or abandon the pattern. Thus, interaction can disturb the fixed regularities of an individual decision process. It can act as a behavioral destructuring mechanism: one player's pattern becomes an object of prediction for the other player, and this prediction pressure can force both players away from their habitual local rules.

This framing is also relevant to the comparison between artificial intelligence and natural intelligence. In artificial systems, randomness is usually supplied by explicitly implemented pseudorandom number generators or by external physical random sources. Natural intelligence is different. Human behavior is generated by an embodied biological system with limited memory, limited motor control, limited attention, and strategic habits. Human randomness therefore provides a concrete behavioral window into natural intelligence. It asks not only how an individual system fails to generate ideal randomness, but also how its limitations can be modified through coupling with another adaptive system. This perspective is consistent with embodied, situated, and distributed views of cognition, according to which intelligent behavior is shaped not only by an isolated internal algorithm, but also by bodily, environmental, and social coupling \cite{Varela1991,Clark1997,Hutchins1995}.

The rock--paper--scissors (RPS) game provides a simple behavioral system for studying this problem. At each time step, a player chooses one of three actions: Rock, Paper, or Scissors. In a one-shot game without prior information, the three moves are symmetric. In repeated play, however, previous moves can be observed, remembered, predicted, and counter-predicted. Thus, repeated RPS is simple enough for controlled analysis, but rich enough to contain human habits, prediction, counter-prediction, and strategic interaction. Repeated RPS has also been used to study bounded rationality, conditional responses, and cyclic behavior in human strategic interaction \cite{WangXuZhou2014}.
More broadly, RPS has been used as a compact model of cyclic dominance and non-equilibrium strategic interaction \cite{Zhou2016RPS}. Recent behavioral studies have also used repeated RPS to formalize opponent modeling and to measure the limits of human adaptive sequential reasoning \cite{BrockbankVul2021OpponentModeling,BrockbankVul2024RepeatedRPS}.

A perfectly random RPS sequence is strategically useful because it prevents the opponent from exploiting regularities. In game-theoretic terms, the equal-probability mixed strategy is the natural equilibrium strategy. However, actual human RPS sequences are not perfectly random.
Our previous work used RPS sequences to extract and estimate human strategic patterns, including characteristic strategy extraction and homology-search-based strategy estimation \cite{KomaiKimKurokawa2018RISP,KomaiKimKousakaKurokawa2019Homology}.
Previous work on Human Randomness in RPS investigated the human capacity to generate randomness in decision-making processes and evaluated RPS time series using Lempel--Ziv complexity and recurrence-plot determinism \cite{KomaiKurokawaKim2022HumanRandomness}. Lempel--Ziv complexity was originally introduced as a complexity measure for finite sequences and is useful here as an operational measure of sequential complexity \cite{LempelZiv1976}. That study showed that human-generated sequences can be quantitatively distinguished from pseudorandom and algorithmically generated sequences. This suggests that human decisions contain structured, biased, or rule-like components even when randomness would be strategically useful.
This focus differs from studies that primarily ask how well players exploit structured opponents; here, we ask whether mutual interaction can also transform the randomness and complexity of the players' own action sequences.

The present study starts from this point and asks a further question: if an individual human system has limitations in generating randomness, to what extent can these limitations be modified through interaction with another human? In other words, we examine human randomness not only as an individual capacity, but also as a possible interaction-dependent property of a coupled system.

The working hypothesis is based on frequency-bias response. Suppose one player uses Rock more frequently than the other moves in a recent time window. The opponent may respond by increasing Paper, because Paper beats Rock. If Paper then becomes frequent, the first player may respond by increasing Scissors. If this process continues, frequency biases may circulate between the two players. If the circulation does not collapse into a simple deterministic cycle, it may reduce persistent frequency bias and move the empirical distribution of moves closer to equal frequency. As a result, entropy may increase. In some cases, the sequence may also become more complex according to Lempel--Ziv complexity.

The idea that simple rules can generate complex or random-looking behavior is not new. In cellular automata, for example, elementary Rule 30 is a well-known case in which a simple local deterministic rule produces patterns that appear random-like \cite{Wolfram1985OriginsRandomness,Wolfram2002}. This does not mean that human interaction is equivalent to a cellular automaton. Rather, Rule 30 provides a useful conceptual example: random-like behavior can arise from repeated local interactions. The present study asks a modest behavioral version of this question: can interaction between two humans enhance the randomness of their action sequences?

We do not attempt to solve the philosophical problem of whether true randomness exists. Nor do we attempt to decide whether physical randomness is fundamental or epistemic. Even coin tossing, often treated as random in practice, can be analyzed as a deterministic physical process under sufficiently precise initial conditions \cite{DiaconisHolmesMontgomery2007}. Physical random number generation also raises practical issues about device bias, detection, and statistical testing. These issues are important, but they are not the main focus of this paper. Here, we use operational measures of randomness and complexity, especially Lempel--Ziv complexity and entropy, to study human behavioral sequences.

Empirically, we analyze human--human RPS data consisting of all pairwise matches among 9 participants. There were 36 unique pairings, and each pairing was repeated 3 times. Each match consisted of 300 consecutive RPS moves by each player. Thus, the dataset contains 108 human--human matches and 216 individual player sequences.

The analysis addresses three questions. First, do human--human RPS sequences produce Lempel--Ziv complexity values that reach or exceed the empirical range observed in the RNG-opponent condition? This tests whether interaction can produce high-complexity individual sequences beyond the range observed when humans played against an RNG-like opponent.

Second, is there a measurable behavioral quantity associated with later entropy increase? We introduce a sensitivity measure that captures whether a player responds to the opponent's recent frequency bias by choosing the move that beats the opponent's most frequent recent move. We then test whether this focal-player sensitivity predicts future entropy in the opponent's move sequence after controlling for the opponent's current entropy.

Third, is the sensitivity--entropy relation specific to temporally aligned human--human interaction? To address this, we compare the observed data with circular-shift surrogates. The surrogate preserves each opponent's sequence structure while disrupting the real-time alignment between the two players. If the observed sensitivity--entropy relation is stronger than the surrogate relation, this suggests that temporally aligned interaction contributes to the effect.


The paper is organized as follows. Section~\ref{sec:methods} defines the experimental setting, the Lempel--Ziv complexity measure, and the sensitivity measure. Section~\ref{sec:results} presents the LZC results, sensitivity analysis, partial regression, and surrogate analysis. Section~\ref{sec:discussion} discusses the interpretation, limitations, and future intervention experiments needed for causal validation. Section~\ref{sec:conclusion} concludes the paper.

\section{Methods}
\label{sec:methods}

This section defines the experimental setting and the measures used in the analysis. We focus on whether human--human interaction can produce RPS sequences with higher operational randomness or complexity than those observed in an RNG-opponent condition.
In the RNG-opponent condition, the maximum observed human LZC value was \(84\). We therefore use \(84\) as an empirical reference value for comparison with the human--human condition. This value is not a theoretical upper bound on human-generated randomness, but only an empirical reference from the RNG-opponent condition.

\subsection{Experimental setting}

In this study, data were collected through a rock–paper–scissors game conducted using a computer. 
Participants selected their moves (rock, paper, or scissors) via keyboard input and played against a computer program.
The computer’s moves were generated using the Python standard random module, which is based on the Mersenne Twister algorithm \cite{MatsumotoNishimura1998}.
Participants were instructed to play with the objective of maximizing their number of wins. 
No information was provided regarding whether the opponent employed a fully random strategy. 
In addition, participants were asked to remain silent, concentrate on the task, and select their moves with the goal of winning. 
As prior information, participants were informed that, if the opponent were completely random, playing randomly would yield the highest winning rate.
The experiment consisted of three sets, each comprising 300 rounds. An interval of at least three hours was imposed between sets to mitigate the effects of sequential trials. 
The time required for data collection depended on the participant’s input speed; however, each set typically took approximately 3 to 5 minutes to complete.

We analyzed human--human RPS data consisting of all pairwise matches among 9 participants.
There were 36 unique pairings, and each pairing was repeated 3 times. Thus, the dataset contains
\[
36 \times 3 = 108
\]
human--human matches. In each match, both players produced a sequence of 300 consecutive RPS moves. Therefore, the total number of individual player series is
\[
108 \times 2 = 216.
\]

Each move was encoded as
\[
0 = \mathrm{Rock}, \qquad
1 = \mathrm{Scissors}, \qquad
2 = \mathrm{Paper}.
\]
The raw data files were named according to the following convention:
\[
\texttt{h[match\_id]-[repeat\_id]\_player[player\_id]},
\]
for example,
\[
\texttt{h24-2\_player1}.
\]

\subsection{Randomness measure: Lempel--Ziv complexity}

As a primary measure of sequence complexity, we used Lempel--Ziv complexity (LZC) \cite{LempelZiv1976}. In this paper, unless otherwise stated, LZC refers to the raw Lempel--Ziv phrase count computed for a sequence of length 300 over the alphabet \(\{0,1,2\}\). This is the same raw LZC measure used in our previous analysis of human-vs-computer RPS sequences.

LZC is used here as an operational measure of sequential complexity. A higher LZC value indicates that the sequence is less compressible by the Lempel--Ziv parsing procedure and therefore contains fewer repeated sequential patterns under this measure. We do not treat LZC as a complete test of mathematical randomness; rather, it is one finite-sequence complexity measure used for comparison across experimental conditions.

\subsection{Sensitivity to the opponent's frequency bias}

We define a sensitivity measure to quantify whether a player responds to the opponent's recent frequency bias in the winning direction. Let \(x_i(t)\in\{0,1,2\}\) denote the move of player \(i\) at time \(t\), where
\[
0=\mathrm{Rock}, \qquad
1=\mathrm{Scissors}, \qquad
2=\mathrm{Paper}.
\]
Let \(j\) denote the opponent of player \(i\). For a window size \(M\), we consider the opponent's recent move history
\[
\{x_j(t-M), x_j(t-M+1), \ldots, x_j(t-1)\}.
\]
Let
\[
b_{j,M}(t)
=
\operatorname{mode}
\{x_j(t-M), x_j(t-M+1), \ldots, x_j(t-1)\}
\]
be the opponent's most frequent move in this window. When there is a tie, we use the most recently observed move among the tied moves as the tie-breaking rule.

Let
\[
w(b_{j,M}(t))
\]
denote the move that beats \(b_{j,M}(t)\). Under our encoding, this winning-response function is
\[
w(b) = (b-1) \bmod 3,
\]
because Rock beats Scissors, Scissors beats Paper, and Paper beats Rock.

We then define the local sensitivity indicator as
\[
S_{i,M}(t)
=
\begin{cases}
1, & \text{if } x_i(t)=w(b_{j,M}(t)),\\
0, & \text{otherwise}.
\end{cases}
\]
Thus, \(S_{i,M}(t)=1\) means that player \(i\) selected the move that beats the opponent's most frequent recent move.

For each individual player series, we compute the time-averaged sensitivity
\[
\bar{S}_{i,M}
=
\frac{1}{T-M}
\sum_{t=M+1}^{T}
S_{i,M}(t),
\]
for \(M=1,\ldots,10\). We then define the maximum sensitivity as
\[
S_{M_o}
=
\max_{M\in\{1,\ldots,10\}}
\bar{S}_{i,M},
\]
where
\[
M_o
=
\arg\max_{M\in\{1,\ldots,10\}}
\bar{S}_{i,M}.
\]
When multiple values of \(M\) give the same maximum value, we choose the smallest \(M\).

The quantity \(S_{M_o}\) is used as a compact series-level measure of how strongly a player tends to respond to the opponent's recent frequency bias in the winning direction.

\begin{figure}[htbp]
    \centering
    \includegraphics[width=0.95\linewidth]{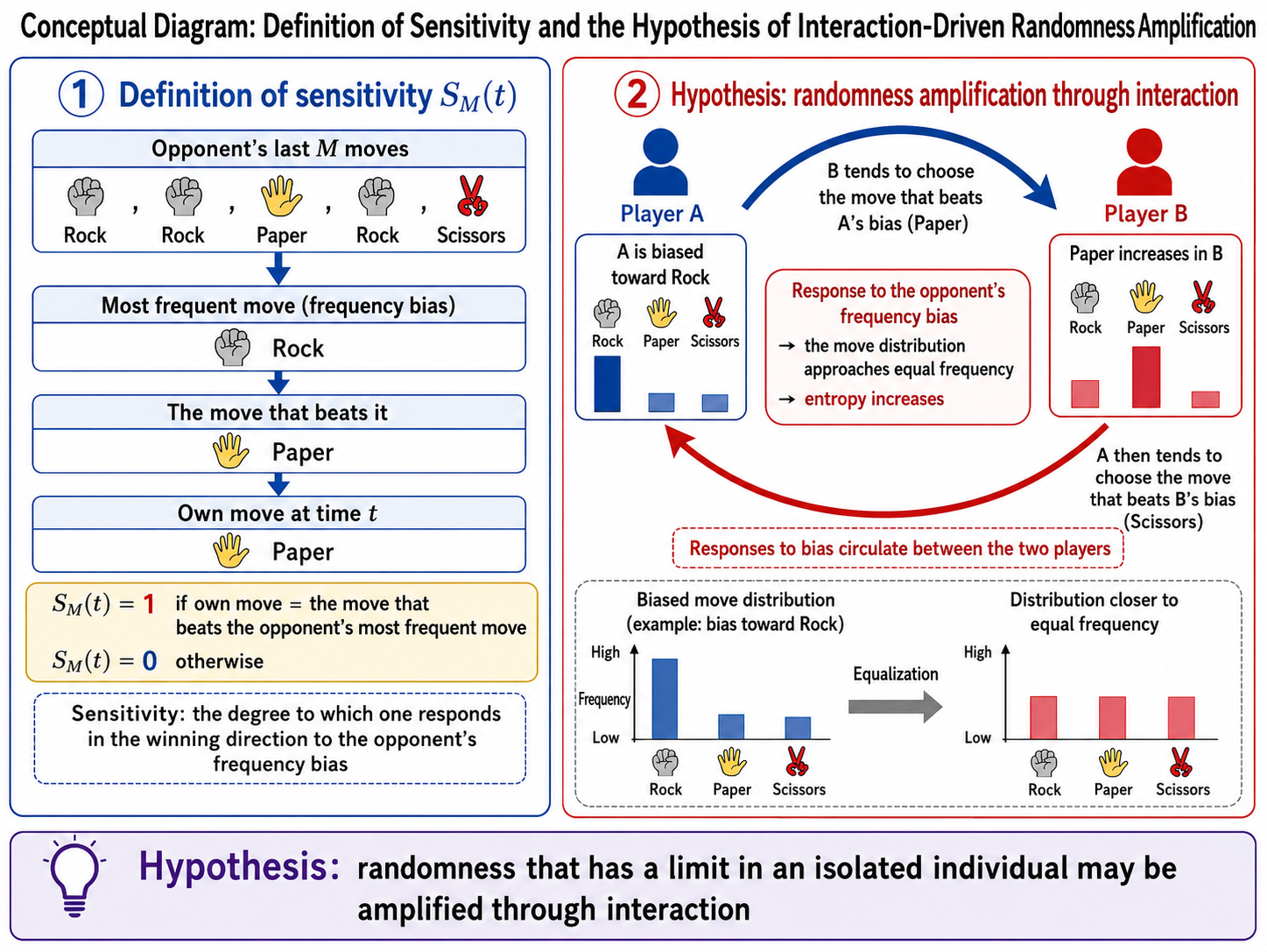}
    \caption{
    Conceptual diagram of sensitivity and the hypothesized interaction-driven randomness amplification.
    The left panel illustrates the definition of the local sensitivity indicator \(S_{i,M}(t)\). The opponent's last \(M\) moves are used to identify the opponent's most frequent recent move, and \(S_{i,M}(t)=1\) if the player's current move is the move that beats that frequency-biased move. The right panel illustrates the working hypothesis that responses to frequency bias can circulate between the two players. If such responses reduce persistent frequency bias, the move distribution may approach equal frequency and entropy may increase. This figure is explanatory and does not by itself constitute evidence for the mechanism.
    }
    \label{fig:sensitivity_definition}
\end{figure}

\subsection{Working hypothesis}

The working hypothesis is as follows.
Suppose one player shows a frequency bias, for example, using Rock more often than the other moves.
The opponent may respond by using Paper more often, because Paper beats Rock. This response can then become a new frequency bias, to which the first player may respond by using Scissors more often.
In this way, frequency biases may circulate between the two players.
In this sense, sensitivity is not defined as a general tendency to change moves, nor as a tendency to maximize immediate winning rate. It is specifically defined as a response in the winning direction to the opponent's recent frequency bias.

If such responses are not locked into a simple deterministic cycle, this interaction may push the empirical distribution of moves closer to equal frequency. As a result, the entropy of the move distribution may increase. This provides a possible mechanism by which human--human interaction can increase the randomness of individual move sequences.

We emphasize that this mechanism is treated as a hypothesis.
The present data first test whether sensitivity is associated with later opponent-side entropy after controlling for current opponent entropy. We then use surrogate analyses to examine whether this association depends on temporal alignment between the two players. Because the proposed mechanism is not expected to operate uniformly across all players and all time points, we also examine a state-conditioned case in which the opponent's recent move distribution is low in entropy and therefore clearly biased. Establishing full intervention-level causality would require additional experiments, such as experimentally manipulating the opponent's frequency bias or instructing participants to respond to frequency bias.

Figure~\ref{fig:sensitivity_definition} summarizes both the operational definition of sensitivity and the hypothesized bias-circulation mechanism. The figure is intended as a schematic guide; the statistical evidence for the mechanism is provided by the following section.

\section{Results}
\label{sec:results}

The results are organized in the following order.
First, we show the reference observation from the RNG-opponent condition, where human LZC reached but did not exceed the empirical reference value of 84.
Second, we analyze LZC in the human--human condition and show that a small number of individual player series exceeded this value.
Third, we examine the sensitivity measure \(S_{M_o}\).
Fourth, we test whether sensitivity predicts later opponent-side entropy using partial regression.
Fifth, we use circular-shift surrogate analyses to test whether the sensitivity--entropy relation is specific to the original temporal alignment.
Because the proposed mechanism is state-dependent rather than uniform, we finally examine a low-entropy opponent state, where the opponent's recent move distribution is biased and the hypothesized sensitivity mechanism should be most visible.

\subsection{Human LZC in the RNG-opponent condition}
\label{subsec:rng_lzc}

\begin{figure}[t]
    \centering
    \includegraphics[width=0.85\linewidth]{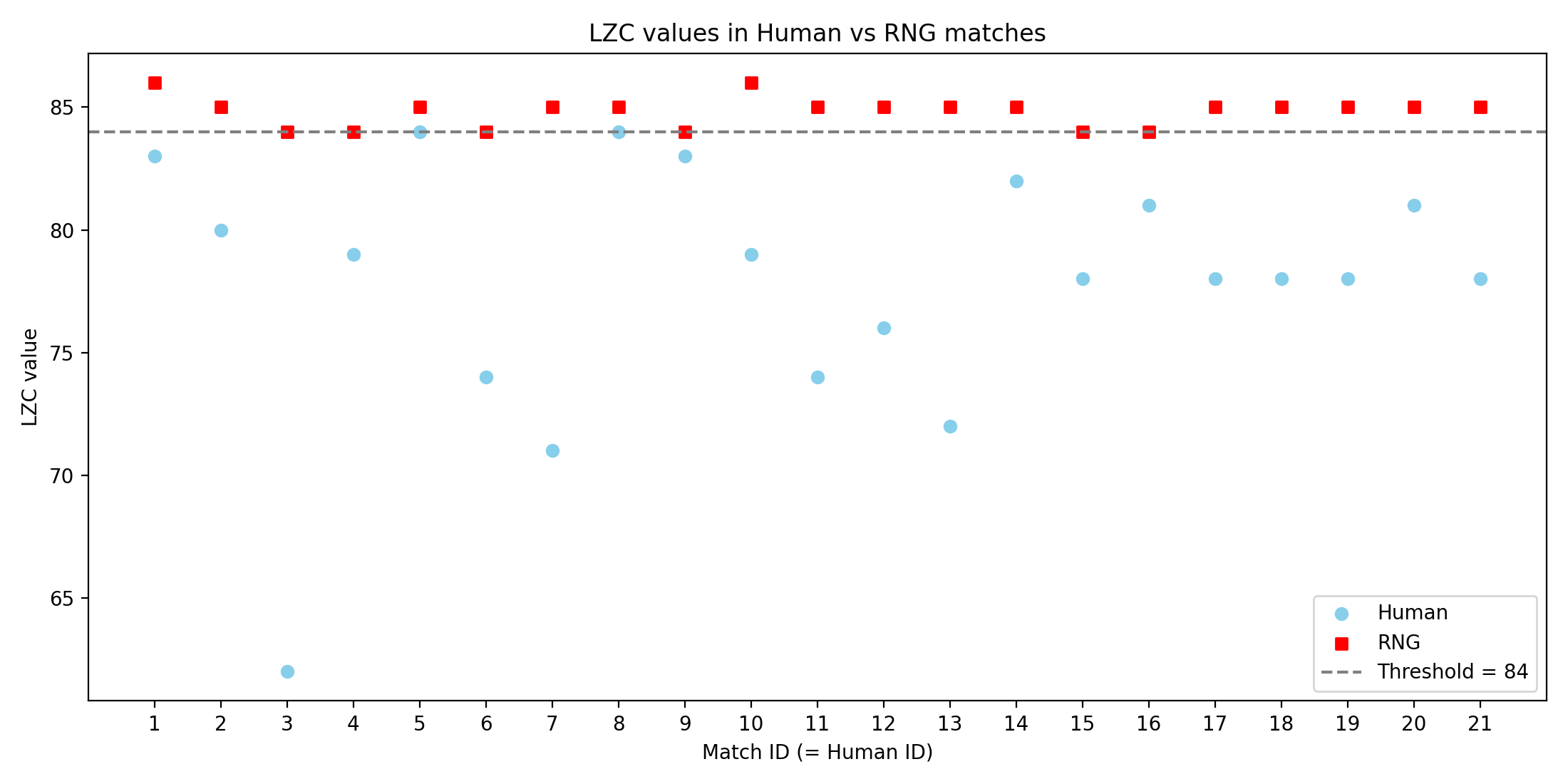}
    \caption{
      Human Lempel--Ziv complexity in the RNG-opponent condition.
      Each point represents the LZC of one human move sequence.
The dashed horizontal line indicates LZC \(=84\), the maximum value observed in the RNG-opponent condition. This value is used as an empirical reference for comparison with the human--human condition.
    }
    \label{fig:rng_lzc}
\end{figure}

Figure~\ref{fig:rng_lzc} shows the LZC of human move sequences in the RNG-opponent condition.
In this condition, the human player played against a computer opponent using the Mersenne Twister pseudorandom number generator \cite{MatsumotoNishimura1998}.
The important observation is that the maximum human LZC value in the RNG-opponent condition was 84; some series reached this value, but none exceeded it. We therefore use LZC \(=84\) as an empirical reference value for comparison with the human--human condition.

This result is used here as a baseline.
It suggests that, when humans generate move sequences against an RNG-like opponent, their LZC remains within a certain empirical range. We do not interpret 84 as a mathematical upper bound.
Instead, we use it as an empirical reference threshold for comparison with the human--human condition.

\subsection{Human--human interaction produced a small number of high-LZC individual series}
\label{subsec:human_human_lzc}

\begin{figure}[htbp]
    \centering
    \includegraphics[width=0.95\linewidth]{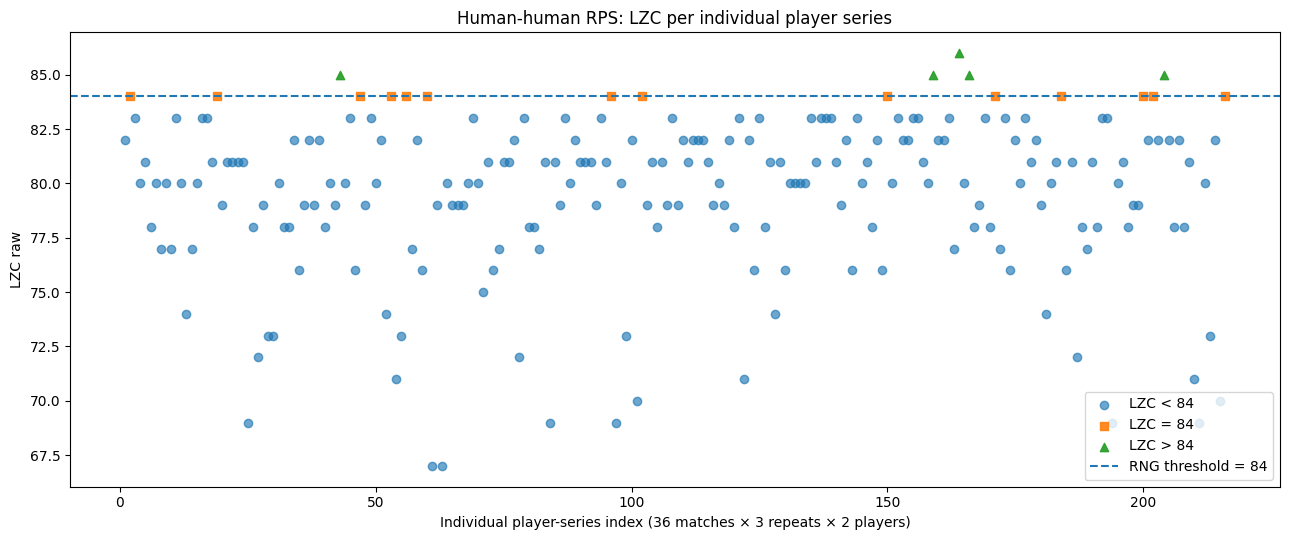}
    \caption{
    LZC of individual player series in the human--human condition.
    The dataset contains 216 individual player series, corresponding to 36 pairings, 3 repetitions, and 2 players per match. The dashed line indicates the empirical RNG threshold, LZC \(=84\). Blue circles indicate series below 84, orange squares indicate series equal to 84, and green triangles indicate series above 84. Most individual series remained below the threshold, but a small number reached or exceeded it. This supports the conservative statement that human--human interaction can produce high-complexity individual sequences in some cases, rather than uniformly increasing LZC across the entire population.
    }
    \label{fig:hh_lzc_scatter}
\end{figure}

We next computed LZC for all 216 individual player series in the human--human condition.
Figure~\ref{fig:hh_lzc_scatter} shows the LZC of each individual series.
The dashed horizontal line again indicates the empirical RNG threshold, LZC \(=84\).

The main observation is that most individual series remained below 84, but some reached or exceeded it.
Specifically, among 216 individual player series, 14 series had LZC equal to 84, 20 series had LZC greater than or equal to 84, and 6 series had LZC strictly greater than 84. The maximum observed LZC was 86.

Thus, the human--human condition did not shift the entire LZC distribution above the RNG threshold. The effect is not a uniform increase across all players and all matches. However, unlike the RNG-opponent condition, the human--human condition produced a small number of individual sequences that reached or exceeded the empirical threshold.

\begin{figure}[htbp]
    \centering
    \includegraphics[width=0.75\linewidth]{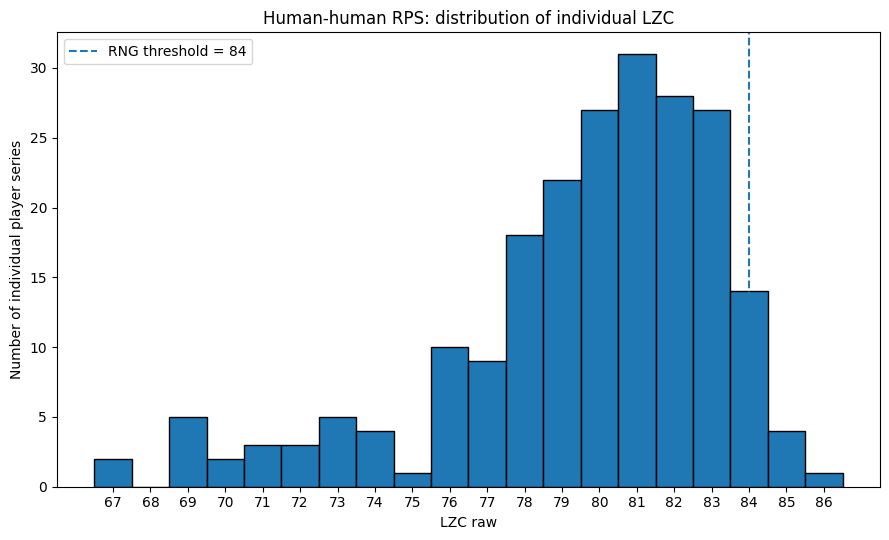}
    \caption{
    Distribution of individual LZC values in the human--human condition.
    The histogram uses integer bins for the raw LZC values. The dashed vertical line indicates the empirical RNG threshold, LZC \(=84\). Among 216 individual player series, 14 series had LZC \(=84\), 20 series had LZC \(\geq 84\), and 6 series had LZC \(>84\). The maximum value was 86. This figure shows that high-LZC cases occur in the right tail of the distribution, not as a general shift of the entire distribution.
    }
    \label{fig:hh_lzc_hist}
\end{figure}

Figure~\ref{fig:hh_lzc_hist} shows the same result as a histogram. The distribution is concentrated around LZC values of approximately 78--83, with a right tail extending to 86. This confirms that the threshold-reaching and threshold-exceeding cases are relatively rare, but they are present.

These results should be interpreted carefully. If we average the two players in each match, the pair mean LZC generally remains below 84. Therefore, the appropriate statement is not that the average human--human match exceeds the RNG threshold. Rather, the appropriate statement is that human--human interaction produced a small number of individual sequences that reached or exceeded the threshold that had not been exceeded in the RNG-opponent condition.

\subsection{Sensitivity to the opponent's frequency bias}
\label{subsec:sensitivity_distribution}

We then examined the series-level maximum sensitivity \(S_{M_o}\), defined in Section~\ref{sec:methods}. This measure summarizes, for each individual player series, how strongly the player tends to choose the move that beats the opponent's most frequent recent move.

\begin{figure}[htbp]
    \centering
    \includegraphics[width=0.95\linewidth]{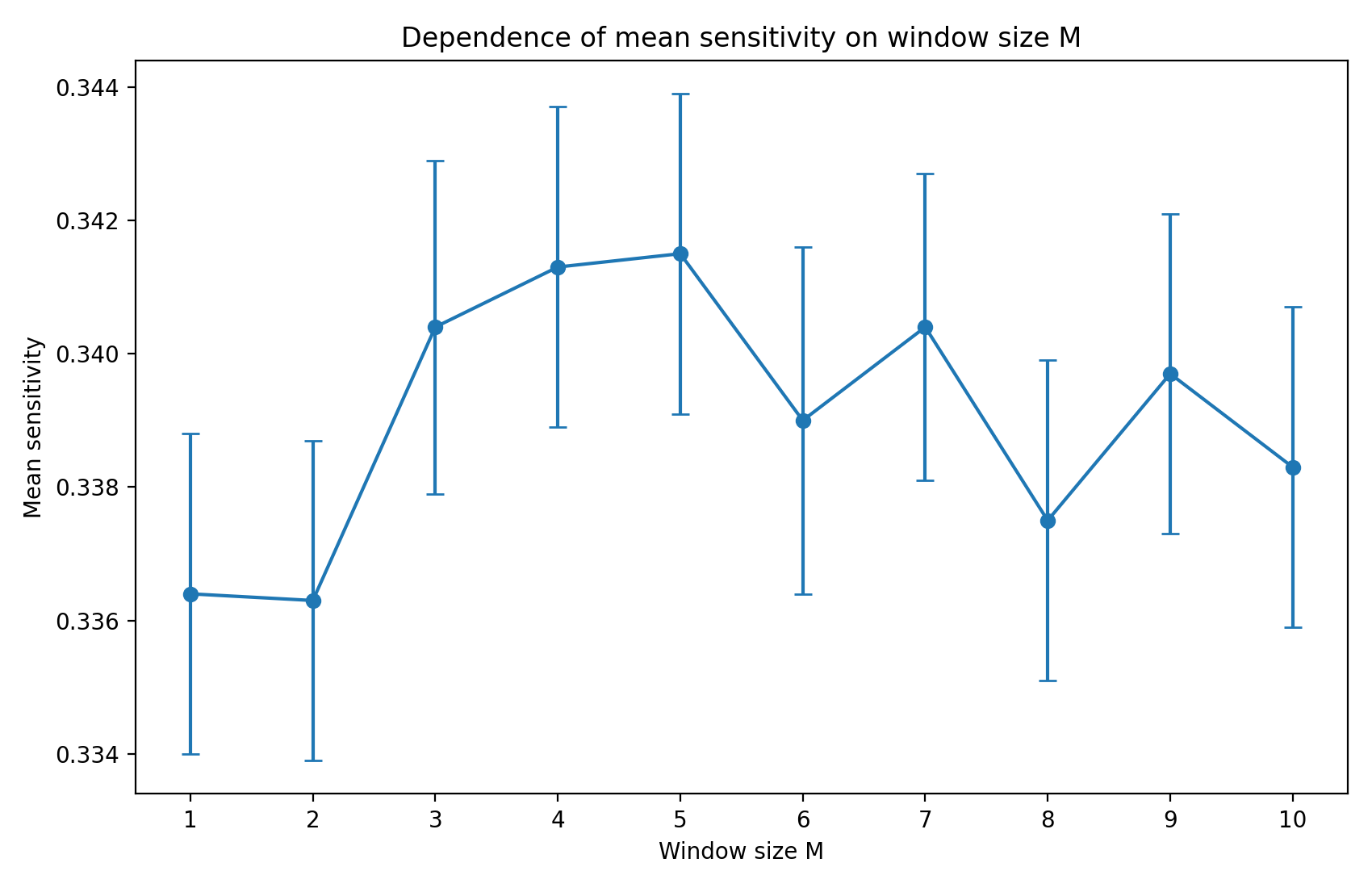}
    \caption{
Dependence of mean sensitivity on the window size \(M\).
Sensitivity was computed for \(M=1,\ldots,10\). Points show the mean across individual player series, and error bars show the standard error of the mean. The chance-level value is \(1/3 \approx 0.333\). The mean sensitivity was slightly above chance for all \(M\), with relatively higher values around \(M=3\)--\(5\). This suggests weak but systematic sensitivity to the opponent's recent frequency bias, without implying a fixed universal memory size.
}
    \label{fig:sensitivity_M_dependence}
\end{figure}

Figure~\ref{fig:sensitivity_M_dependence} summarizes the dependence of mean sensitivity on the window size \(M\). Across \(M=1,\ldots,10\), the mean sensitivity stayed in a narrow range, approximately 0.336--0.342. All mean values were slightly above the chance-level value \(1/3 \approx 0.333\), which would be expected if the player's current move were independent of the opponent's recent frequency bias. This suggests that the sensitivity measure captures a weak but systematic tendency to respond to the opponent's recent bias in the winning direction.

The dependence on \(M\) was not large, but the mean sensitivity was relatively higher around \(M=3,4,5\). This may suggest that players respond not only to the immediately preceding move, but also to a short recent history of the opponent's moves. However, the differences across \(M\) were small, and we do not interpret this as evidence for a fixed universal memory size. Rather, the result indicates that sensitivity is present across several short window sizes, with a possible weak preference for short memory windows around \(M=3\)--\(5\).

\begin{figure}[htbp]
    \centering
    \includegraphics[width=0.95\linewidth]{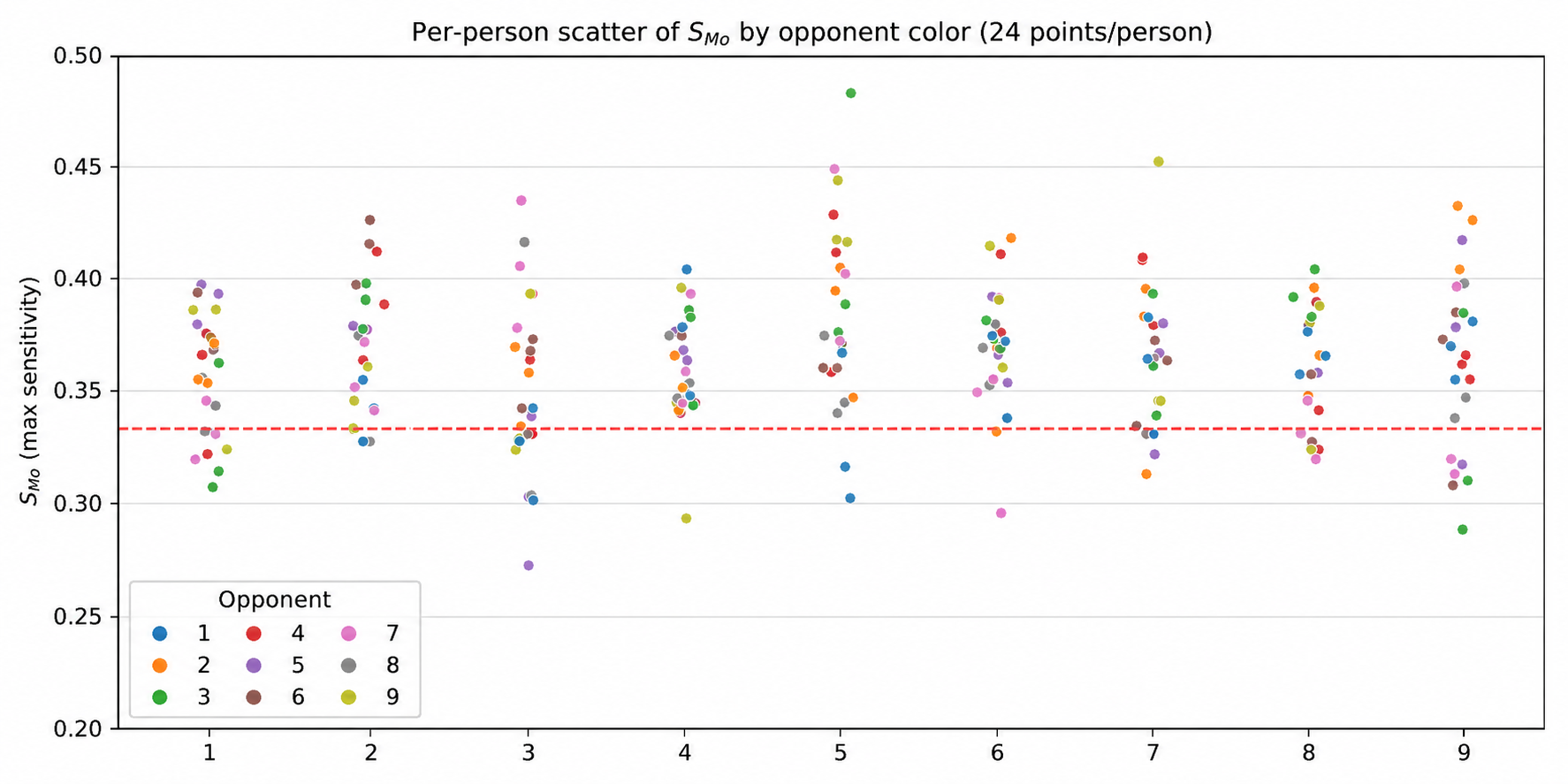}
      \caption{
Maximum sensitivity \(S_{M_o}\) by anonymized player identifier.
For each player series, sensitivity was computed across multiple window sizes \(M\), and the maximum value \(S_{M_o}\) was used as a series-level sensitivity score. The figure shows that sensitivity differs across players. These differences should be interpreted cautiously: the present data do not allow us to separate fixed individual traits from opponent-dependent strategic behavior.
}
    \label{fig:sensitivity_max_by_person}
\end{figure}

Figure~\ref{fig:sensitivity_max_by_person} summarizes the maximum sensitivity \(S_{M_o}\) by player. This figure is used to show that sensitivity varies across individuals. However, we do not interpret these differences as stable personality traits. They may reflect individual strategy, the opponent, the particular match, and the strategic state during the sequence.

\begin{figure}[htbp]
    \centering
    \includegraphics[width=0.75\linewidth]{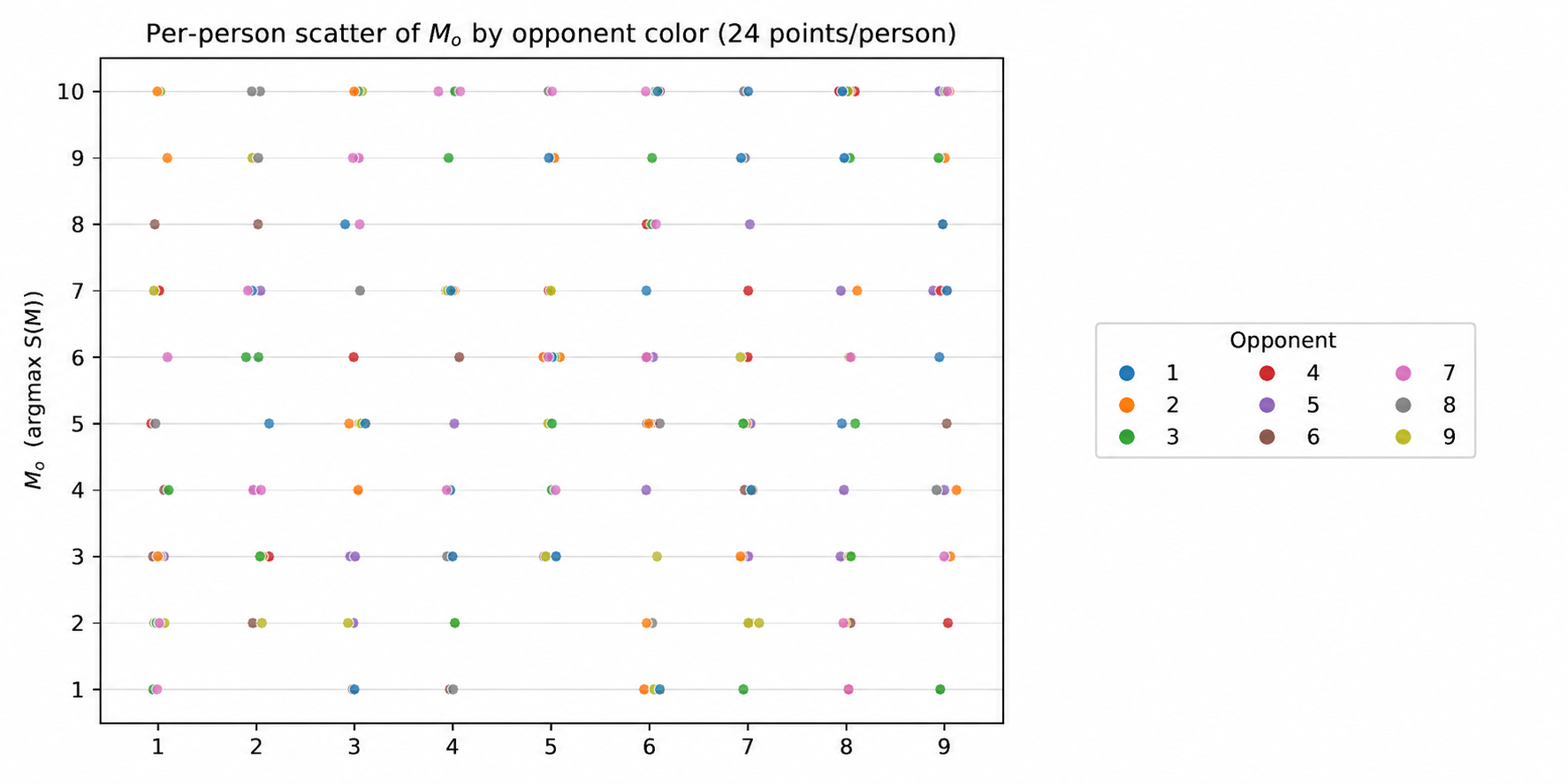}
    \caption{
Memory size of maximum sensitivity \(M_o\) by anonymized player identifier across opponent identifiers.
For each individual player series, \(M_o\) is the window size \(M\in\{1,\ldots,10\}\) that gives the maximum mean sensitivity. The figure plots these \(M_o\) values separately for each anonymized player and opponent pairing. The distribution of \(M_o\) values is heterogeneous across players. Some players show a relatively concentrated range of \(M_o\) values, whereas others show a broader distribution across several values of \(M\). However, even players with an apparent preference for a particular \(M_o\) still exhibit variation across opponents and matches. Therefore, the figure does not support the existence of a single universal memory size shared by all players. Instead, it suggests flexible and interaction-dependent use of multiple short memory windows.
}
    \label{fig:sensitivity_max_M_dependence}
\end{figure}

Figure~\ref{fig:sensitivity_max_M_dependence} summarizes the memory size of maximum sensitivity, \(M_o\), for each player across opponents. Here, \(M_o\) denotes the window size \(M\) that gives the maximum mean sensitivity for each individual player series. The figure shows substantial heterogeneity across players. Some players appear to use a relatively concentrated range of \(M_o\) values, suggesting a preference for particular short memory sizes, whereas others show a broader spread across many values of \(M_o\). At the same time, even players who appear to favor a particular \(M_o\) do not use only one memory size; their \(M_o\) values still vary across opponents and matches. Thus, the result does not support a single fixed universal memory size. Rather, it suggests that players differ in how narrowly or broadly they rely on recent opponent history, while still showing flexible use of multiple short memory windows depending on the interaction.

\FloatBarrier
\subsection{Sensitivity predicts later entropy increase}
\label{subsec:partial_regression}

We next tested whether the focal player's sensitivity is associated with later entropy increase in the opponent's move sequence. Let \(H_j(t)\) denote the Shannon entropy of opponent \(j\)'s recent move distribution at time \(t\), computed over an entropy window of length \(M_{\mathrm{ENT}}\). Following the analysis design, we define the future entropy change as
\[
\Delta H_j(t)
=
\frac{H_j(t+1)+H_j(t+2)+H_j(t+3)}{3}
-
H_j(t).
\]
This quantity measures whether the opponent's local move entropy increases over the next three time steps relative to the current entropy. This formulation tests an interaction-level relation: whether player \(i\)'s sensitivity to player \(j\)'s recent frequency bias is associated with subsequent entropy increase in player \(j\)'s move sequence.

The basic question is whether
\[
S_i(t)
\]
is positively associated with
\[
\Delta H_j(t),
\]
where \(S_i(t)\) indicates whether player \(i\) responds to player \(j\)'s recent frequency bias in the winning direction. However, current opponent entropy \(H_j(t)\) must be controlled for, because a low-entropy state has more room to increase than an already high-entropy state. Therefore, we used the following partial regression model:
\[
\frac{H_j(t+1)+H_j(t+2)+H_j(t+3)}{3}
=
\beta_0
+
\beta_1 S_i(t)
+
\beta_2 H_j(t)
+
\varepsilon(t).
\]
The key coefficient is \(\beta_1\). A positive \(\beta_1\) means that, after controlling for the opponent's current entropy level, higher focal-player sensitivity predicts higher future entropy in the opponent's move sequence. Equivalently, it supports a positive relation between focal-player sensitivity and subsequent opponent-side entropy increase.

Table~\ref{tab:partial_regression} summarizes the pooled partial regression results. The coefficient of focal-player sensitivity was positive for both entropy-window settings. For \(M_{\mathrm{ENT}}=10\), the sensitivity coefficient was 0.01898 with \(p=7.81\times10^{-18}\). For \(M_{\mathrm{ENT}}=5\), the coefficient was 0.04601 with \(p=7.73\times10^{-22}\). Thus, focal-player sensitivity positively predicted future entropy in the opponent's move sequence after controlling for the opponent's current entropy.

\begin{table}[htbp]
  \centering
  \caption{
  Pooled partial regression of opponent-side future entropy on focal-player sensitivity while controlling for current opponent entropy.
  The dependent variable is the opponent's average future entropy
  \(\{H_j(t+1)+H_j(t+2)+H_j(t+3)\}/3\).
  The explanatory variables are focal-player sensitivity \(S_i(t)\) and the opponent's current entropy \(H_j(t)\).
  Cluster-robust standard errors were computed using individual player series as clusters.
  The table reports results for two entropy-window settings, \(M_{\mathrm{ENT}}=5\) and \(M_{\mathrm{ENT}}=10\).
  }
  \label{tab:partial_regression}
  \begin{tabular}{lccccc}
    \hline
    Entropy window & \(n_{\mathrm{series}}\) & \(n_{\mathrm{points}}\) & \(\beta_1\) & SE & \(p\)-value \\
    \hline
    \(M_{\mathrm{ENT}}=5\)  & 216 & 62,746 & 0.04601 & 0.00479 & \(7.73\times 10^{-22}\) \\
    \(M_{\mathrm{ENT}}=10\) & 216 & 61,992 & 0.01898 & 0.00221 & \(7.81\times 10^{-18}\) \\
    \hline
  \end{tabular}
\end{table}
The positive coefficient should be interpreted as an interaction-level association. It does not mean that a player's sensitivity immediately increases the entropy of the player's own sequence. Rather, it indicates that when player \(i\) responds to player \(j\)'s recent frequency bias in the winning direction, player \(j\)'s subsequent move distribution tends to become more entropic, after controlling for player \(j\)'s current entropy. This is consistent with the proposed bias-circulation mechanism, in which one player's response changes the strategic state faced by the other player.

Although the numerical values of the sensitivity coefficients may appear small in absolute terms, they should be interpreted on the entropy scale of a three-action system. The maximum possible entropy is only \(\log_2 3 \approx 1.585\) bits, and the regression concerns a local change over the next three time steps rather than a whole-sequence transformation. Therefore, coefficients of this magnitude are not negligible in the present setting; they indicate a small but meaningful local shift in opponent-side entropy.

These results should not be interpreted as proof of a causal effect. The regression shows temporal prediction and statistical association between focal-player sensitivity and subsequent opponent-side entropy, but it does not by itself show that the association is specific to the original real-time interaction between the two players. We therefore next examine whether the sensitivity--entropy relation survives surrogate comparisons, first across all time points and then in a low-entropy opponent state.

\FloatBarrier

\subsection{All-series circular-shift surrogate analysis}
\label{subsec:surrogate_all}

To test whether the sensitivity--entropy relation depends on real-time interaction, we constructed circular-shift surrogate data. Surrogate-data methods are commonly used to test whether an observed time-series statistic depends on a specific temporal structure rather than on simpler properties of the series \cite{Theiler1992}. In each surrogate, the opponent sequence was circularly shifted. This operation preserves the opponent's move sequence, including its frequency distribution and much of its serial structure, but disrupts the original real-time alignment between the two players.

The surrogate analysis asks whether the sensitivity--future entropy relation is stronger in the original temporally aligned data than in shifted alignments. If the same relation appears after circular shifting, then the effect may be explained by individual sequence structure rather than by real-time interaction timing. If the observed data show a stronger effect than the surrogate data, this supports the interpretation that temporal alignment contributes to the effect.

We first performed an all-series surrogate analysis using the same opponent-side entropy model as in the partial regression. For each individual player series, we used the series-specific \(M_o\) obtained from the sensitivity analysis and computed the same regression coefficient as in the observed data. We used \(K\)-gap circular shifts with \(K=5,10,20\), excluding near-zero shifts so that short-lag temporal coupling was not included in the surrogate baseline.

The all-series surrogate analyses did not show a robust observed--surrogate advantage. For \(K=10\), the mean observed-minus-surrogate difference across all 216 individual series was close to zero, and the positive fraction was only slightly above one half. Similar results were obtained for \(K=5\) and \(K=20\). Thus, when all eligible time points and all individual series were averaged together, the temporal-alignment-specific component of the sensitivity--entropy relation was weak.

This result is informative rather than simply negative. The proposed sensitivity mechanism is not expected to operate uniformly across all players, dyads, and time points. If the opponent's current move distribution is already high in entropy, there is little room for further entropy increase.
This motivates a state-conditioned surrogate analysis focused on low-entropy opponent states.

\subsection{State-conditioned surrogate analysis in low-entropy opponent states}
\label{subsec:surrogate_low_entropy}

The working hypothesis predicts that sensitivity should be most relevant when the opponent's recent move distribution is biased. We therefore performed a state-conditioned surrogate analysis restricted to low-entropy opponent states. Specifically, using \(M_{\mathrm{ENT}}=10\), we retained only time points satisfying
\[
H_j(t)\leq 1.295.
\]
This condition selects time points where the opponent's recent move distribution is relatively far from equal frequency and therefore has room to move toward higher entropy.

We then fitted the same opponent-side regression model as before,
\[
\frac{H_j(t+1)+H_j(t+2)+H_j(t+3)}{3}
=
\beta_0
+
\beta_1 S_i(t)
+
\beta_2 H_j(t)
+
\varepsilon(t),
\]
using only the low-entropy time points. We compared the observed pooled coefficient \(\beta_1\) with \(K=10\) gap-shift circular surrogates. The gap-shift surrogate used circular shifts \(k=10,\ldots,T-10\), thereby excluding near-zero positive and negative shifts.

In the low-entropy state-conditioned analysis, the pooled observed sensitivity coefficient was larger than the mean coefficient obtained from the \(K=10\) gap-shift surrogate distribution. The observed-minus-surrogate difference was positive, and the empirical right-tail probability relative to the surrogate distribution was \(p=0.00709\). Table~\ref{tab:low_entropy_surrogate} summarizes the analysis.

\begin{table}[htbp]
    \centering
    \caption{
    Low-entropy state-conditioned surrogate analysis.
    Only time points satisfying \(H_j(t)\leq 1.295\) were retained, using \(M_{\mathrm{ENT}}=10\). The observed pooled sensitivity coefficient was compared with \(K=10\) gap-shift circular surrogates. The result shows that, in low-entropy opponent states, the temporally aligned data had a stronger sensitivity--future entropy relation than the shifted surrogate baseline.
    }
    \label{tab:low_entropy_surrogate}
    \begin{tabular}{lc}
        \hline
        Quantity & Value \\
        \hline
        Entropy condition & \(H_j(t)\leq 1.295\) \\
        Entropy window \(M_{\mathrm{ENT}}\) & 10 \\
        Future horizon \(H\) & 3 \\
        Gap-shift parameter \(K\) & 10 \\
        Pooled observed points & 7466 \\
        Observed coefficient \(\beta_{\mathrm{obs}}\) & 0.014407 \\
        Mean surrogate coefficient \(\overline{\beta}_{\mathrm{surr}}\) & 0.001801 \\
        Observed-minus-surrogate difference & 0.012606 \\
        \(z\) versus surrogate distribution & 2.726 \\
        Empirical right-tail probability & 0.00709 \\
        Observed pooled regression \(p\)-value & 0.0295 \\
        \hline
    \end{tabular}
\end{table}

The corresponding series-level analysis was directionally positive but not statistically reliable. This is expected because the low-entropy restriction left relatively few time points per individual series. The mean number of observed low-entropy points per valid series was much smaller than in the full regression. Therefore, the state-conditioned result should be interpreted primarily as pooled evidence for a low-entropy interaction state, not as a uniform effect across all individual series.

These results support a state-dependent interpretation of the sensitivity mechanism. When the opponent's recent behavior was already biased, the original temporally aligned data showed a stronger sensitivity--future entropy relation than the gap-shift surrogate data. This suggests that sensitivity can contribute to a local transition from biased move frequencies toward higher entropy.

\FloatBarrier
\subsection{Summary of the results}
\label{subsec:results_summary}

The results can be summarized as follows.

First, in the RNG-opponent condition, human LZC did not exceed the empirical reference value of 84. This supports the idea that human-generated RPS sequences have a limited empirical range in this non-human opponent condition.

Second, in the human--human condition, most individual LZC values remained below 84, but 20 out of 216 individual series reached or exceeded 84, and 6 exceeded 84. Thus, human--human interaction produced a small high-complexity tail that was not observed in the RNG-opponent condition. This effect was not a uniform shift of the whole LZC distribution.

Third, focal-player sensitivity to the opponent's recent frequency bias was positively associated with later entropy in the opponent's move sequence. This association remained after controlling for the opponent's current entropy and was robust across the two entropy-window settings \(M_{\mathrm{ENT}}=5\) and \(M_{\mathrm{ENT}}=10\).

Fourth, all-series circular-shift surrogate analyses did not show a robust temporal-alignment-specific effect when all time points were averaged together. This indicates that the sensitivity--entropy association is not uniformly expressed across the entire dataset.

Fifth, the state-conditioned surrogate analysis revealed a clearer effect in low-entropy opponent states. When the opponent's recent move distribution was biased, defined by \(H_j(t)\leq 1.295\), the pooled observed sensitivity coefficient was larger than the \(K=10\) gap-shift surrogate coefficient. This supports the interpretation that sensitivity contributes to entropy increase primarily in states where the opponent's recent behavior contains a clear frequency bias.

Taken together, the results support a state-dependent version of interaction-driven randomization. Human--human interaction does not uniformly make all sequences random. However, it can create local conditions in which a biased move distribution is destabilized through temporally aligned response to the opponent's frequency bias. At the same time, this local entropy-increase mechanism should not be interpreted as a complete explanation of the high-LZC tail. LZC is a sequence-level measure, and exceeding the empirical reference value may require longer-term strategic changes, mutual prediction, chains of responses and counter-responses, or multiple interacting mechanisms.

\section{Discussion}
\label{sec:discussion}

This study examined whether human--human interaction can enhance the randomness of human-generated RPS sequences. The main result is not that human interaction uniformly produces random behavior. Rather, the results show a more limited and state-dependent pattern. In the RNG-opponent condition, human LZC did not exceed the empirical reference value of 84. In the human--human condition, most individual sequences also remained below this value, but a small number exceeded it.

In addition, focal-player sensitivity to the opponent's recent frequency bias predicted later entropy in the opponent's move sequence after controlling for the opponent's current entropy. All-series surrogate analyses did not isolate a robust temporal-alignment-specific effect across all time points. However, when the analysis was restricted to low-entropy opponent states, where the opponent's recent behavior contained a clear frequency bias, the temporally aligned observed data showed a stronger sensitivity--future entropy relation than gap-shift circular surrogates.

Taken together, these results support a cautious interpretation: human--human interaction can contribute to randomness enhancement in a state-dependent manner. The evidence identifies a local mechanism by which interaction may destabilize biased behavior and increase entropy, but it does not prove a complete causal mechanism or explain all high-LZC sequences.

\subsection{Interpretation: state-dependent interaction-driven randomization}
\label{subsec:discussion_mechanism}

The proposed interpretation is that human--human interaction can increase randomness through responses to frequency bias, but only under appropriate strategic states. If one player uses one move more frequently, the opponent may increase the move that beats that biased move. This response may then become a new bias, to which the first player can respond. In this way, frequency biases may circulate between the two players.

A key point is that individual constraints are not only limitations; in interaction, they can also become targets for the opponent. A habitual bias can be detected, exploited, and thereby destabilized. In this sense, the very regularities that limit an isolated individual's randomness may become leverage points for interaction-driven change.

The low-entropy state-conditioned result clarifies this mechanism. When the opponent's recent move distribution is already high in entropy, there is little room for further entropy increase. In contrast, when the opponent is in a low-entropy state, the recent distribution contains a clearer frequency bias. In such states, a sensitivity-like response has a meaningful target and can help move the subsequent distribution toward higher entropy.

The present findings therefore suggest a simple local route toward interaction-driven randomization. A biased local distribution can be destabilized through a temporally aligned response to the opponent's recent frequency bias. This does not imply that all sensitivity responses are randomizing. A rigid response may produce a predictable cycle or reinforce a pattern. Rather, the result indicates that, in low-entropy opponent states, the observed temporal alignment between the two players carries information relevant to the sensitivity--future entropy relation.

This mechanism should be understood as local and partial. It does not by itself explain why some individual sequences reached or exceeded the empirical LZC reference value. LZC measures sequence-level complexity over the entire 300-move series, whereas the sensitivity analysis captures a local relation between a player's response and the opponent's near-future entropy. Therefore, exceeding the LZC reference value may require more than local entropy increase. It may involve longer-term strategic changes, mutual prediction, chains of responses and counter-responses, or the combination of multiple mechanisms. The sensitivity mechanism identified here should therefore be regarded as one elementary route by which interaction can destabilize local bias, not as a complete account of high-complexity sequence generation.

A more direct test of this mechanism would require additional analyses or new experiments. For example, one could explicitly estimate whether an opponent's Rock bias predicts an increase in the player's Paper use, whether an opponent's Paper bias predicts an increase in Scissors, and whether an opponent's Scissors bias predicts an increase in Rock. Such an analysis would test the proposed bias-response mechanism more directly. The present paper focuses on establishing a first empirical basis for the effect using LZC, partial regression, and state-conditioned surrogate analysis.

\subsection{Relation to generative mechanisms of randomness}
\label{subsec:discussion_rule30}

The present interpretation is related to a broader question: how can random-like behavior be generated from non-random rules? A well-known example is elementary cellular automaton Rule 30. For binary cell states, Rule 30 updates the state of cell \(i\) from the states of the left neighbor, the cell itself, and the right neighbor. One algebraic representation of the rule is
\[
s_i(t+1)
=
s_{i-1}(t)
+
s_i(t)
+
s_{i+1}(t)
+
s_i(t) \cdot s_{i+1}(t)
\pmod{2}.
\]
Thus, a simple deterministic local rule can generate complex and apparently random patterns when iterated over time. In particular, sampling a fixed cell over time, or using suitable subsequences, can produce pseudorandom sequences. Rule 30 has been studied by Wolfram as a simple source of random-like behavior and has also been used in random-number generation in the Wolfram System \cite{Wolfram1985OriginsRandomness,Wolfram2002,WolframRandomNumberGeneration}.

We do not claim that human--human RPS interaction implements Rule 30 or any specific cellular automaton. The analogy is more limited. Rule 30 illustrates that randomness-like sequences can emerge from deterministic local interactions. The present study asks whether an analogous, but behavioral, phenomenon may occur in human interaction. In our setting, the local interaction rule is not a fixed binary cellular-automaton rule. It is a noisy and heterogeneous strategic response: a player may respond to the opponent's recent frequency bias by choosing the move that beats the biased move.

This difference is important. In Rule 30, the update rule is explicit, deterministic, and applied uniformly across all cells and time steps. In human--human RPS, the ``rule'' is neither fixed nor perfectly consistent. Human strategies change across individuals, opponents, and time. Therefore, the expected empirical signature is not a clean deterministic pattern, but a weak population-level effect.
This is broadly consistent with what we observe: sensitivity predicts later opponent-side entropy in the pooled regression, but the temporal-alignment-specific effect becomes clearest only after conditioning on low-entropy opponent states. Thus, the empirical signature is not a clean uniform rule, but a weak and state-dependent behavioral effect.

This comparison suggests a possible future direction. One could construct generative models of RPS interaction in which each player updates their move probabilities according to the opponent's recent frequency bias. Such models would not be intended as exact psychological models. Rather, they would provide candidate causal interaction rules that can be compared with the observed LZC, entropy, sensitivity, and surrogate results. If a simple interaction rule can reproduce the main empirical signatures, it would provide a more explicit generative account of interaction-driven randomness enhancement.

The three-state structure of RPS may also be relevant. Unlike binary cellular automata, RPS has three actions and a cyclic dominance relation: Rock beats Scissors, Scissors beats Paper, and Paper beats Rock. This cyclic structure may allow two interacting players to generate variability through mutual response to bias. However, the present data do not establish the conditions under which such variability emerges. Future work should compare the present empirical results with generative models that vary the number of states, the memory window, the strength of response to bias, and the amount of stochastic noise. Such comparisons may help clarify when interaction produces predictable cycles and when it produces randomness-like behavior.

\subsection{Heterogeneity of interaction effects}
\label{subsec:discussion_heterogeneity}

The present results show substantial heterogeneity. Human--human interaction did not shift the entire LZC distribution above the empirical RNG reference value. Instead, only a small number of individual series reached or exceeded that value. Similarly, the all-series surrogate analyses did not show a robust temporal-alignment-specific effect across all time points. The clearest surrogate-supported effect appeared only after conditioning on low-entropy opponent states.

This heterogeneity is not surprising for human strategic behavior. In repeated RPS, players may change strategies, react differently to different opponents, or fall into temporary cycles. A response to the opponent's frequency bias may increase entropy in one strategic state, but may produce a more predictable pattern in another. Sensitivity is therefore not automatically randomizing. It can destabilize bias when the opponent's recent behavior is biased, but it may also reinforce cycles when both players respond rigidly.
This interpretation is consistent with previous RPS studies showing that human sequential behavior is heterogeneous and cannot be reduced to a single rule such as win-stay/lose-change \cite{ZhangMoisanGonzalez2021RPS}. Large-scale RPS data also indicate that players use opponent history strategically, especially when such information is useful for improving performance \cite{BatzilisEtAl2019MillionRPS}.

This heterogeneity should be viewed as part of the phenomenon rather than as a failure of the analysis. The main claim is not that every human--human interaction increases randomness. The claim is that interaction can create local states in which biased behavior is destabilized and entropy increases. The low-entropy surrogate result provides evidence for such a state-dependent mechanism, while the LZC result suggests that high-complexity sequences may arise only in a subset of interactions.

\subsection{Scope of the present evidence}
\label{subsec:discussion_scope}

The partial regression shows that focal-player sensitivity predicts later entropy in the opponent's move sequence after controlling for the opponent's current entropy. The low-entropy surrogate analysis further shows that this relation is stronger in temporally aligned data than in gap-shift surrogate data when the opponent's recent behavior is biased.

At the same time, the present evidence should be interpreted within its proper scope. The analysis does not establish intervention-level causality. Unobserved factors such as attention, fatigue, local winning or losing streaks, or strategic switching may influence both sensitivity and entropy change. Therefore, we interpret the results as evidence consistent with a state-dependent interaction mechanism, rather than as a complete causal proof.

The empirical reference value LZC \(=84\) should also be interpreted as a reference from the RNG-opponent condition, not as a universal boundary of human randomness. The important observation is comparative: the RNG-opponent condition did not produce human sequences above this value, whereas the human--human condition produced a small number of such sequences. However, the low-entropy sensitivity mechanism does not fully explain the emergence of the high-LZC tail. Larger datasets, intervention experiments, and generative models are needed to determine how local entropy-increase mechanisms contribute to whole-sequence complexity.

\subsection{Future directions: causal and generative tests}
\label{subsec:discussion_future}

The present study focuses on a small set of analyses that directly address the main empirical question: LZC for sequence-level complexity, partial regression for sensitivity-related entropy increase, and circular-shift surrogates for temporal interaction specificity.
Future work can build on this basis by testing more detailed causal and generative models.

A direct causal test would require intervention experiments. One possible design is a human-vs-computer experiment in which the computer opponent deliberately changes its frequency bias across blocks. For example, in one block the computer could use Rock more frequently, in another block Scissors more frequently, and in another block Paper more frequently. If human players respond by increasing the move that beats the induced bias, and if this response leads to later entropy increase, the causal role of frequency-bias response would be more strongly supported.

A second design is an instruction-based intervention. Participants could be randomly assigned to different instruction conditions. In one condition, they would be instructed to observe the opponent's recent frequency bias and respond with the move that beats it. In another condition, they would be instructed to generate moves as randomly as possible. In a third condition, they would receive no special instruction. If the sensitivity-instruction condition produces larger entropy increase or higher LZC, this would provide stronger evidence that sensitivity contributes to randomness amplification.

A third direction is to construct generative models of RPS interaction. Such models could specify how each player updates move probabilities in response to the opponent's recent frequency bias. The models could then be compared with the observed LZC distribution, sensitivity coefficients, entropy changes, and surrogate results. This would provide a more explicit bridge between the present empirical findings and a mechanistic account of interaction-driven randomness enhancement.

Finally, future analyses could test the proposed bias-circulation mechanism more directly.
For example, if the opponent is biased toward Rock, the model should test whether the player subsequently increases Paper; if the opponent is biased toward Paper, whether the player increases Scissors; and if the opponent is biased toward Scissors, whether the player increases Rock. Distance from the uniform distribution, such as \(D_{\mathrm{KL}}(p_t\|u)\), could also be used to test whether sensitive responses move the empirical distribution closer to equal frequency.
Such an intervention would allow the induced bias, the participant's sensitivity response, and the subsequent entropy change to be temporally separated and tested within the same experimental design.

A further direction is to model the emergence of the high-LZC tail.
The low-entropy state-conditioned analysis identifies a local mechanism: when the opponent's recent behavior is biased, sensitivity to that bias is associated with higher future opponent entropy, and this relation is stronger in temporally aligned data than in gap-shift surrogates. However, LZC is not a purely local entropy measure. It reflects how repeated patterns accumulate or fail to accumulate over the entire sequence. Therefore, a sequence exceeding the empirical LZC reference value may require additional processes beyond local movement toward equal frequencies.

Future generative models should therefore examine longer-range interaction dynamics, including strategic switching across phases of a match, mutual prediction and counter-prediction, response chains between players, and transitions between different behavioral modes. Such models could test whether local sensitivity alone is sufficient to reproduce the observed LZC distribution, or whether multiple interacting mechanisms are needed to generate the highest-complexity sequences.

Relatedly, recent work on game-driven random walks has shown that strategies based on an opponent's hand history can generate correlated game outcomes, and that mutual adaptive strategies may produce persistent fluctuations rather than convergence to a fixed equilibrium \cite{Tsuji2025GameDrivenRandomWalks}. Although that model is not directly comparable to human RPS behavior, it supports the view that high-LZC sequences may require longer-range strategic adaptation and response chains beyond local entropy increase alone.
This view is also related to previous work on random walks with time-dependent bias, where periodic, quasi-periodic, or chaotic bias sequences can strongly affect diffusion behavior \cite{KimNaruseAonoHoriAkimoto2016ChaoticBias}. In the present behavioral setting, the effective bias is not externally imposed but may be generated endogenously through interaction between two players.

A related future direction is to examine self-side entropy as a possible second-order feedback effect.
The present main analysis focuses on whether focal-player sensitivity predicts later entropy in the opponent's move sequence. However, a cyclic interaction account also predicts that the focal player's own later entropy may change after the opponent responds to the focal player's action.
Exploratory self-side surrogate analyses showed a globally positive observed--surrogate tendency, and the observed--surrogate difference was better explained by directed player--opponent grouping than by focal player identity alone.
This suggests that self-side entropy effects may depend on who is responding to whom rather than on fixed individual traits.
Because this interpretation involves an additional interaction step, we leave it for future work on longer-range interaction dynamics and high-LZC sequence generation.

\subsection{Implications for natural intelligence}

The results suggest that human randomness should not be treated only as an individual ability. This is consistent with embodied, situated, and distributed views of natural intelligence, in which behavior is shaped by the coupling between an agent, its body, its environment, and other agents \cite{Varela1991,Clark1997,Hutchins1995}.
A person may have limited ability to generate random-like sequences in isolation or against a non-human opponent. However, interaction with another human may create a dynamic environment in which frequency biases are detected, responded to, and modified. In this sense, randomness can be partly interaction-driven.

The use of entropy as a behavioral index in competitive interaction is also consistent with neurobehavioral work showing that, in a matching-pennies game, anterior insula activity tracked participants' own behavioral entropy rather than the opponent's entropy \cite{TakahashiEtAl2015AnteriorInsulaEntropy}. This suggests that behavioral entropy can be a meaningful variable for studying strategic adjustment, although the present study focuses on interaction-level entropy dynamics rather than neural correlates.

This does not mean that interaction always improves randomness. It means that interaction can create conditions under which higher-complexity sequences emerge. The important point is that randomness here may be a property of a coupled system, not only a property of an isolated individual.

\section{Conclusion}
\label{sec:conclusion}

This study investigated whether human--human interaction can enhance randomness in the rock--paper--scissors game. We analyzed 108 human--human matches, corresponding to 216 individual player sequences of 300 moves each.

The results support a cautious but meaningful conclusion. In the RNG-opponent condition, human LZC did not exceed the empirical reference value of 84. In the human--human condition, most individual sequences remained below this value, but 20 out of 216 sequences reached or exceeded it, and 6 sequences exceeded it. Thus, human--human interaction did not uniformly increase LZC, but it produced a small high-complexity tail not observed in the RNG-opponent condition.

We then examined sensitivity to the opponent's recent frequency bias. Sensitivity was defined as choosing the move that beats the opponent's most frequent recent move. Partial regression showed that focal-player sensitivity positively predicted future entropy in the opponent's move sequence after controlling for the opponent's current entropy. This result was robust across the two entropy-window settings \(M_{\mathrm{ENT}}=5\) and \(M_{\mathrm{ENT}}=10\).

Surrogate analyses clarified the scope of this result. When all time points were averaged together, circular-shift surrogate analyses did not isolate a robust temporal-alignment-specific effect. However, when the analysis was restricted to low-entropy opponent states, where the opponent's recent move distribution was clearly biased, the pooled observed sensitivity coefficient was larger than the coefficient obtained from \(K=10\) gap-shift circular surrogates. This supports a state-dependent interpretation: sensitivity can contribute to entropy increase when the opponent's behavior contains a clear frequency bias.

The present study therefore supports the hypothesis that human--human interaction can shape randomness, but not as a uniform effect across all players and all time points. The evidence points to a local mechanism in which temporally aligned responses to frequency bias can destabilize a biased move distribution and move it toward higher entropy. At the same time, this mechanism does not fully explain why some sequences exceeded the empirical LZC reference value. High-LZC sequence generation may require longer-term strategic changes, mutual prediction, response chains, or multiple interacting mechanisms.

A more precise causal test will require intervention experiments that manipulate the opponent's frequency bias, the participant's sensitivity strategy, or the available feedback. Generative models will also be needed to determine whether local sensitivity mechanisms are sufficient to reproduce the high-LZC tail. The present findings provide a first empirical basis for treating human randomness not only as an individual property, but also as a state-dependent property of interaction between natural intelligent systems.

\bmhead{Acknowledgements}

The author used ChatGPT (OpenAI) for English editing and takes full responsibility for the final version.

\section*{Declarations}

\paragraph{Funding}
This work was supported by SOBIN Institute LLC under Research Grant SP011.

\paragraph{Competing interests}
The authors declare that there are no competing interests.

\paragraph{Ethics approval and consent to participate}
The study analyzed low-risk behavioral data from a rock--paper--scissors task.
Participant names were used internally to link repeated matches and, where necessary, subsequent strategy-related interviews, but participant identities were anonymized or pseudonymized in the manuscript, figures, and data summaries.
No sensitive personal information was collected or analyzed.
Participants provided consent for the use of their behavioral data for research purposes.
Ethics approval was not required under the applicable institutional rules for this type of low-risk anonymized behavioral study.

\paragraph{Consent for publication}
Not applicable.

\paragraph{Data availability}
The datasets generated and analyzed during the current study contain participant-linked behavioral records.
De-identified data may be made available from the corresponding author on reasonable request, subject to participant consent and applicable institutional rules.


\paragraph{Code availability}
The analysis code used during the current study is available from the corresponding author on reasonable request.


\paragraph{Materials availability}
Not applicable.

\paragraph{Author Contributions}
S.-J. K. and H. K. conceived the research theme. S.-J. K. defined the specific research direction and methodology, implemented and executed the code, obtained the results, and drafted the manuscript. S. O. and H. K. collected the data used in this study. All authors discussed the results and contributed to revising and improving the manuscript.

\begin{appendices}

  \FloatBarrier
  \section{Supplementary surrogate analyses}
\label{app:surrogate_supplement}

This appendix summarizes additional surrogate analyses used to evaluate the robustness and scope of the sensitivity--entropy relation. The main text reports the low-entropy state-conditioned pooled surrogate result because it most directly tests the proposed mechanism: sensitivity should matter most when the opponent's recent move distribution is biased. Here we report additional analyses showing that the effect is not uniformly present across all time points or across simple \(S_{M_o}\)-defined subgroups.

  \subsection{All-series gap-shift surrogate analysis}

We first repeated the circular-shift surrogate analysis across all individual series and all eligible time points using \(K\)-gap shifts with \(K=5,10,20\). Across these settings, the mean observed-minus-surrogate difference was close to zero, and the positive fraction was only slightly above one half. Thus, the all-series surrogate analysis did not provide robust evidence for a uniform temporal-alignment-specific effect.

Table~\ref{tab:appendix_all_series_gap_surrogate} summarizes the all-series gap-shift surrogate analyses.

\begin{table}[htbp]
    \centering
    \caption{
    All-series gap-shift circular surrogate analyses.
    For each individual player series, the observed sensitivity coefficient was compared with the mean coefficient obtained from gap-shift circular surrogates. The gap parameter \(K\) excludes near-zero positive and negative shifts. \(\Delta\beta\) denotes the observed-minus-surrogate difference,
    \(\Delta\beta=\beta_{\mathrm{obs}}-\overline{\beta}_{\mathrm{surr}}\).
    Across \(K=5,10,20\), the mean difference was close to zero, and the positive fraction was only slightly above one half. Thus, the all-series surrogate analysis did not provide robust evidence for a uniform temporal-alignment-specific effect.
    }
    \label{tab:appendix_all_series_gap_surrogate}
    \begin{tabular}{lcccccc}
        \hline
        Gap \(K\) & \(n\) & Mean \(\Delta\beta\) & Median \(\Delta\beta\) & Positive fraction & \(t\)-test \(p\) & Wilcoxon \(p\) \\
        \hline
        5  & 216 & -0.000120 & 0.000896 & 0.537 & 0.900 & 0.503 \\
        10 & 216 & -0.000080 & 0.001007 & 0.551 & 0.934 & 0.500 \\
        20 & 216 & -0.000059 & 0.000918 & 0.546 & 0.951 & 0.496 \\
        \hline
    \end{tabular}
\end{table}

\subsection{\(S_{M_o}\)-defined subgroup analyses}

We also examined whether the observed-minus-surrogate difference was larger in high-\(S_{M_o}\) series. The results did not show a robust monotonic strengthening in high-\(S_{M_o}\) groups. The highest-\(S_{M_o}\) group showed a slightly positive median difference and a positive fraction above one half, but the mean difference was close to zero and the tests were not significant. This suggests that high sensitivity alone does not necessarily imply entropy-enhancing interaction. Sensitivity may reflect either flexible bias-destabilizing behavior or rigid response patterns, depending on the strategic state.

\subsection{Median-split entropy-state analysis}

We next divided time points into low- and high-entropy states according to the opponent's current entropy within each panel. This analysis showed weak directional support: the mean observed-minus-surrogate difference was positive in the low-entropy state and negative in the high-entropy state. However, the median differences and positive fractions were close to zero, and the tests were not significant. This motivated the more targeted low-entropy-only analysis reported in the main text.

\subsection{Low-entropy-only pooled analysis}

Finally, we restricted the analysis to time points satisfying \(H_j(t)\leq 1.295\). Because this restriction leaves relatively few time points per individual series, series-level estimates were noisy. We therefore also performed a pooled regression across all retained low-entropy time points, using series-clustered standard errors. This pooled analysis produced a stronger observed coefficient than the \(K=10\) gap-shift surrogate baseline, as reported in the main text.

\end{appendices}


\clearpage
\bibliography{sn-bibliography}

\end{document}